\documentclass[conference]{IEEEtran}
\IEEEoverridecommandlockouts
\let\svtilde~

\newcommand\newtildeON[1][A]{\def~{\csname newtilde#1\endcsname}}
\newcommand\newtildeOFF{\let~\svtilde}
\usepackage{cite}
\makeatletter 
\newcommand{\linebreakand}{%
  \end{@IEEEauthorhalign}
  \hfill\mbox{}\par
  \mbox{}\hfill\begin{@IEEEauthorhalign}
}
\makeatother 

\usepackage{amsmath,amssymb,amsfonts}
\usepackage{algorithmic}
\usepackage{graphicx}
\usepackage[bottom]{footmisc}
\usepackage{textcomp}
\usepackage{listings}
\usepackage[table]{xcolor}
\usepackage{mathtools}
\usepackage{diagbox}
\usepackage{enumitem} 
\usepackage{subfigure}
\usepackage{graphicx}
\definecolor{codegreen}{rgb}{0,0.6,0}
\definecolor{codegray}{rgb}{0.5,0.5,0.5}
\definecolor{codepurple}{rgb}{0.58,0,0.82}
\definecolor{backcolour}{rgb}{0.95,0.95,0.92}

\definecolor{dkgreen}{rgb}{0,0.6,0}
\definecolor{gray}{rgb}{0.5,0.5,0.5}
\definecolor{mauve}{rgb}{0.58,0,0.82}

\def\BibTeX{{\rm B\kern-.05em{\sc i\kern-.025em b}\kern-.08em
    T\kern-.1667em\lower.7ex\hbox{E}\kern-.125emX}}
\begin{document}

\title{Fighting Game Adaptive Background Music for Improved Gameplay}

\author{
\IEEEauthorblockN{Ibrahim Khan$^1$, Thai Van Nguyen$^1$, Chollakorn Nimpattanavong$^1$, Ruck Thawonmas$^2$}
\IEEEauthorblockA{
\{$^1$Graduate School, $^2$College\} of Information Science and Engineering, Ritsumeikan University, Japan\\
\{gr0556vx,gr0557fv, gr0608sp\}@ed.ritsumei.ac.jp, ruck@is.ritsumei.ac.jp
}}
\IEEEoverridecommandlockouts

\IEEEpubid{\makebox[\columnwidth]{979-8-3503-2277-4/23/\$31.00~\copyright2023 IEEE \hfill}
\hspace{\columnsep}\makebox[\columnwidth]{ }}
\maketitle
\IEEEpubidadjcol

\begin{abstract}
This paper presents our work to enhance the background music (BGM) in DareFightingICE by adding adaptive features. The adaptive BGM consists of three different categories of instruments playing the BGM of the winner sound design from the 2022 DareFightingICE Competition. The BGM adapts by changing the volume of each category of instruments. Each category is connected to a different element of the game. We then run experiments to evaluate the adaptive BGM by using a deep reinforcement learning AI agent that only uses audio as input (Blind DL AI). The results show that the performance of the Blind DL AI improves while playing with the adaptive BGM as compared to playing without the adaptive BGM.
\end{abstract}


\section{Introduction}

This research focuses on video game music in fighting games and how to use it to enhance players’ experience. In fighting games, background music (BGM) should change in response to the dynamic action, creating suspense and excitement as the players engage in combat. \par
Our work is based on the DareFightingICE Competition, which has two tracks: Sound Design Track and AI Track \cite{b2}. The competition has run at the  IEEE Conference on Games since 2022. The platform for this competition is DareFightingICE -- an enhanced version of FightingICE \cite{b3}. \par



The goal of this research is to create an adaptive BGM by modifying the BGM of DareFightingICE, and evaluate the performance of the adaptive BGM by using a deep learning AI agent that only uses audio as input (Blind DL AI) \cite{b4}. Our research focuses on giving players information about the state of the game through the BGM. We are also the first group to use multiple instruments' volumes in the adaptation of a BGM.

\section{Related Work}
Since our research touches on the topic of adaptive fighting game music, we describe a few fighting games that have adaptive BGM.\par

Adaptive or dynamic, as it is called in commercial fighting games, music has been used in a few fighting games in the past. The most prominent examples are Killer Instinct, Tekken 7, and Them's Fightin' Herds. For Takken 7, many states have a more intense version of the BGM being played for the last round. Them's Fightin' Herds mixes in bits of each character's themes depending on who's winning. Lastly, Killer Instinct has multiple layers to its adaptive BGM that changes according to the movement of the players as well as the actions of the players. \par

While adaptive BGM exists in fighting games, it is used to enhance the overall gaming experience of the players, however, currently, no BGM gives information to players such as the distance between players, and the amount of health points (HP) remaining for both players. It is also worth noting that adaptive music exists in games outside of fighting games as well, but since our focus is on fighting games, we only discussed adaptive music in this genre.\par




\section{Rule-Based Adaptive BGM}
\begin{figure}[t!]
\centerline{\includegraphics[width=0.45\textwidth]{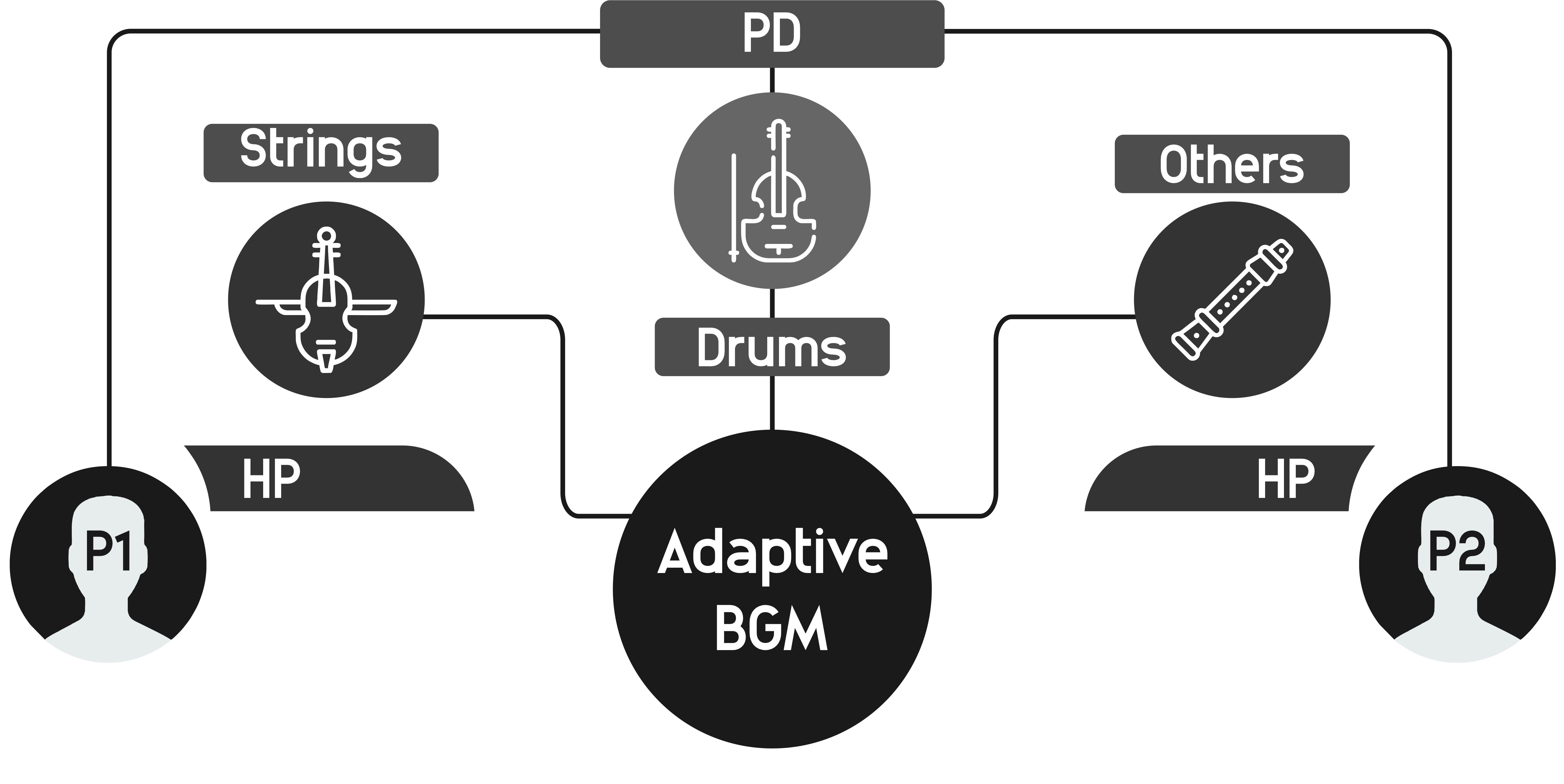}}
\caption{Outline of Adaptive BGM.}
\label{figADM}
\end{figure}

In this research, we propose an adaptive BGM that adapts to player actions and the players' in-game positions. The proposed adaptive BGM consists of three different categories of instruments playing the BGM available in the winning sound design from the 2022 DareFightingICE competition. The BGM's instruments were separated using an AI service called Moises \cite{b20}. The resulting three different instrument groups are the Drums, the Strings, and the Others category available after the separation. The BGM adjusts by altering the loudness of the instruments. The game elements in use are both players’ HP and the distance between the two players (PD). The HP is the number of hits a player can take before losing. The design of the proposed adaptive BGM is illustrated in Fig.~\ref{figADM}. \par

In Fig.~\ref{figADM}, the strings are connected to player 1’s HP, and the others are connected to player 2’s HP. Lastly, the drums are connected to PD.\par

The maximum HP for one player per round is 400, and the PD is 800 pixels horizontally. The HP of both players and their PD are divided into different levels to change the volumes of the different groups of instruments gradually. The levels are empirically selected due to the lack of existing research when it comes to adaptive BGM using volume modulation.\par

The levels for HP are 400, 300, 250, 200, 150, 100, and 50, which are connected to the respective instrument group's volume levels of 75\%, 60\%, 55\%, 40\%, 35\%, 25\%, and 10\% respectively. For PD, the levels are 800, 600, 500, 400, 300, 60, and 0,  which are connected to the drums' volume levels of 10\%, 20\%, 30\%, 40\%, 50\%, 60\%, and 75\%.

The adaptive BGM is created in DareFightingICE. Since there has been a recent update to the DareFightingICE platform from version 5.2 (used in the 2022 competition) to version 6.0, we use the latest version.\par




\section{Experiments}
For evaluation of the adaptive BGM, we conducted an objective evaluation, the details of which are as follows. \par

We use the aforementioned Blind DL AI. Our protocol for this evaluation is to train the Blind DL AI with and without the adaptive BGM and then compare the performances of the AI agents. We hypothesize that if the performance of the Blind DL AI is better with the adaptive BGM then the adaptive BGM conveys useful information. we did not compare our approach with previous approaches since they do not exist when it comes to fighting games.\par

We trained the Blind DL AI for 900 rounds on DareFightingICE (1 game has 3 rounds) against MCTSAI23i, a sample agent in the 2023 competition using Monte-Carlo tree search \cite{b18}. The same approach was used in the 2022 competition to train the Blind DL AI.\par

We used similar environments as in the 2022 DareFightingICE Competition. More specifically, six computers were used that have the same specification, i.e., CPU: Intel(R) Xeon(R) W-2135 CPU@ 3.70GHz 3.70 GHz, RAM: 16 GB, GPU: NVIDIA Quadro P1000 4GB VRAM, and OS: Windows 10.\par

\subsection{Results}

Since the Blind DL AI had three different audio encoders \cite{b3} as an option, the experiments were run on each encoder for both versions, i.e., one-dimensional Convolutional Neural Network (1D-CNN), Fast Fourier Transform (FFT), and Mel-Spectrogram (Mel-Spec).\par

We evaluated the performance of each trained Blind DL AI by making it fight against the aforementioned opponent AI agent for 90 rounds. The ratio of the number of wins\footnote{In the game, the round winner is either the one with a non-zero HP while its opponent's HP has reached zero or the one with the higher HP when the round-length limit of 60 s has reached.} over 90 rounds ($n = 90$), Eqn. \eqref{equWinRate}, and the average HP difference at the end of round $r$ between the trained AI agent and its opponent, Eqn. \eqref{HPDifference}, are then calculated. The equations and details above are taken from previous work \cite{b4}. \par
\begin{equation}
    win_{ratio} = \frac {\textit{winning rounds}}{\textit{n}}\label{equWinRate}
\end{equation}
\begin{equation}
    avg HP_{diff} = \frac  { \sum_{r=1}^{r=n}  (HP_r^{self} - HP_r^{opp})}{\textit{n}} \label{HPDifference}
\end{equation}

\begin{table}[t!]
\caption{Performance of the blind DL AI with winner sound design of the 2022 DareFightingICE competition in DareFighingICE 6.0.}
\label{tblFightResults6}
\begin{center}
\begin{tabular}{|c|c|c|c|}
\hline
Sound design &Encoder & $win_{ratio}$ & $avg HP_{diff}$\\
\hline
Winner &1D-CNN & 0.54 & 12.66\\
\hline
Winner &FFT & 0.52 & 27.03\\
\hline
Winner &Mel-spectrogram &  0.56 &  34.23\\
\hline
Winner Adaptive &1D-CNN & {\bf 0.81} & {\bf 101.17}\\
\hline
Winner Adaptive &FFT &{\bf 0.94} & {\bf 136.14}\\
\hline
Winner Adaptive &Mel-spectrogram &{\bf  0.68} &{\bf  74.56}\\
\hline
\end{tabular}
\end{center}
\end{table}

As shown from the results in Table \ref{tblFightResults6}, the Blind DL AI performs better with the adaptive BGM. For each encoder, Blind DL AI also performs better with our adaptive BGM. This proves our hypothesis that the adaptive BGM provides useful information with FFT being the best encoder. Example gameplay videos of blind DL AI with each audio encoder, where the blind DL AI is player 1 and MCTSAI23i is player 2, as well as the separated audio file of each group of instruments and the modified code for the adaptive BGM, can be found on this link\footnote{Link to videos and other files: https://tinyurl.com/aBGM-CoG2023}. On the day of the presentation we will ask for volunteers from the audience to test our adaptive BGM.  \par


\section{Conclusions And Future Work}
This paper presented a rule-based adaptive background music (BGM) system that consists of three different categories of instruments playing the BGM of the winner sound design from the 2022 DareFightingICE Competition. The proposed adaptive BGM adapts by changing the volumes of the different categories of instruments. Each category is connected to a different element of the game. The paper also showed that the performance of a deep reinforcement learning AI agent using only audio data as its input, called Blind DL AI, improved while playing with the adaptive BGM as compared to playing without it.\par

\end{document}